\documentclass[12pt,english,aps,manuscript]{article}
\usepackage[T1]{fontenc}
\usepackage[latin1]{inputenc}
\usepackage{geometry}
\usepackage{amsmath}
\usepackage{amssymb}
\usepackage[numbers]{natbib}
\PassOptionsToPackage{normalem}{ulem}
\usepackage{ulem}

\makeatletter

\usepackage{geometry}

\makeatletter

\usepackage{color}

\makeatletter
\newcommand{\bee}{\begin{equation}}
\newcommand{\ee}{\end{equation}}
\newcommand{\beea}{\begin{eqnarray}}
\newcommand{\eea}{\end{eqnarray}}

\makeatother

\makeatother

\makeatother

\usepackage{babel}
\begin{document}
COLO-HEP-577

\begin{center}
\textbf{\Large A Local Evaluation of Global Issues in SUSY breaking}{\Large{} }
\par\end{center}{\Large \par}

\begin{center}
\vspace{0.3cm}
 
\par\end{center}

\begin{center}
{\large S. P. de Alwis$^{\dagger}$ } 
\par\end{center}

\begin{center}
Abdus Salam International Centre for Theoretical Physics, Strada Costiera
11, \\
 Trieste 34014, Italy 
\par\end{center}

\begin{center}
and
\par\end{center}

\begin{center}
Physics Department, University of Colorado, \\
 Boulder, CO 80309 USA 
\par\end{center}

\begin{center}
\vspace{0.3cm}

\par\end{center}

\begin{center}
\textbf{Abstract} 
\par\end{center}

It is well known that there are different global (i.e. $M_{P}\rightarrow\infty$)
limits of $N=1$ supergravity. We distinguish between these limits
and their relevance to low energy phenomenology. We discuss a) fermion
mass matrices and recently proved theorems in global SUSY b) stability
issues and SUSY breaking d) R-symmetry and a recently derived bound
on the superpotential and e) FI terms in global and local SUSY. 

\vfill{}

$^{\dagger}$ {\small e-mail: dealwiss@colorado.edu}{\small \par}

\eject

\section{Introduction}

There are two categories of theories which attempt to describe supersymmetry
breaking phenomena at the TeV scale%
\footnote{For reviews see for example \citep{Baer:2006rs,Drees:2004jm}.%
}. One is the class of GMSB theories which typically have a relatively
low scale SUSY breaking, with the effective F-term of the SUSY breaking
field(s) at scales around $10^{5}GeV$. The other consists of various
gravity mediated theories in which the corresponding scale lies in
a range $10^{10}GeV-10^{12}GeV$. These correspond to a very low mass
( $\lesssim1eV$) gravitino in the first class and to very heavy $>1TeV$
ones in the latter case. 

The literature on SUSY breaking seems to have split accordingly into
two cultures. The first, the GMSB and dynamical SUSY breaking literature,
is almost exclusively confined to global SUSY. While the necessity
of canceling the cosmological constant (CC) is of course recognized
- it is generally viewed as something to done at the end of the day
by adding a constant to the superpotential. This of course is meaningful
only in the context of supergravity since in global SUSY a constant
in the superpotential is not of any physical significance as it disappears
from the action. On the other hand the literature on the second -
gravity mediation - class of models naturally has to be concerned
with the full supergravity action from the beginning. 

The point of this note is to evaluate various arguments made in the
global SUSY breaking literature (for a recent review see for example
\citep{Kitano:2010fa}) from a local SUSY point of view . In the next
section we define various relevant gravity decoupling limits and discuss
some general issues concerning arguments derived in the context of
global SUSY. In section \ref{sec:sGoldstino-mass} we argue that the
scalar partner of the Goldstino - which is effectively the lightest
neutral scalar in the theory - cannot be raised arbitrarily far above
the scale of the gravitino. In section \ref{sec:R-symmetry-and-SUSY}
we discuss some issues related to R-symmetry and its breaking in global
and local SUSY. The final topic is concerned with the relation between
Fayet-Illiopoulos terms in global and local supersymmetry, and may
serve to clarify some aspects of the discussion of these issues initiated
in \citep{Komargodski:2009pc,Komargodski:2010rb} (see also \citep{Dienes:2009td}).

\section{Local to global }

The action for supergravity at the two derivative level%
\footnote{We choose $M_{P}\equiv(8\pi G_{Newton})^{-1/2}=1$ except where for
clarity it is explicitly written out. Note the dimensions (mass) $[\phi]=1,\,[K]=2,\,[W]=3$.%
} is uniquely determined in terms of the Kaehler potential $K=K(\phi,\bar{\phi})$
the (holomorphic) superpotential $W=W(\phi)$, and gauge coupling
functions $f_{a}(\phi)$ for each simple (or $U(1)$) factor group,
where $ $$\phi\equiv\{\phi^{i}\},\,\phi^{i}=(\phi^{i},\chi^{i},F^{i})$
is the set of chiral scalar fields in the theory. It is often convenient
to define also the combination $G\equiv K+\ln|W|^{2}$ in which case
at a generic point in field space (i.e. away from $W=0$) the action
can be written in terms of $G$ and $f$. The scalar potential is
then given by the formula
\begin{eqnarray}
V & = & e^{G}(G_{i}G^{i}-3)+V_{D}\label{eq:V}\\
 & = & e^{K/M_{P}^{2}}(D_{i}WK^{i\bar{j}}D_{\bar{j}}\bar{W}-3\frac{|W|^{2}}{M_{P}^{2}})+V_{D}\label{V2}
\end{eqnarray}
$G_{i}=\partial_{i}G,\, G_{i\bar{j}}=\partial_{i}\partial_{\bar{j}}G=K_{i\bar{j}},\, G^{i}=G^{i\bar{j}}G_{\bar{j}}$,
and $V_{D}$ is the D-term potential. In the second line we have for
future reference explicitly indicated the scaling with respect to
the Planck mass. The Kaehler covariant derivative (on scalars) is
defined as $D_{i}W=\partial_{i}W+K_{i}W/M_{P}^{2}$. The squared gravitino
mass is given by $m_{3/2}^{2}=e^{G}=e^{K/M_{P}^{2}}|W|^{2}/M_{P}^{4}$
when evaluated at a minimum of the potential.

It should be emphasized that these formulae depend only on the existence
of a two derivative supersymmetric action. It is also important to
stress that quantum effects are not expected to violate SUSY and will
only change the functional form of these functions. Also according
to the SUSY non-renormalization theorems W will not get perturbative
corrections while $f$ will only get such corrections at one loop
though both can have non-perturbative corrections. 

There are several important gravity decoupling limits: 
\begin{eqnarray}
{\rm a})\, M_{P}\rightarrow\infty & \frac{\phi}{M_{P}}\rightarrow0 & \frac{K}{M_{P}^{2}},\,\frac{W}{M_{P}}\rightarrow0\label{eq:globala}\\
{\rm a'})\, M_{P}\rightarrow\infty & \frac{\phi}{M_{P}}\rightarrow0 & \frac{K}{M_{P}^{2}}\rightarrow0,\,\frac{W}{M_{P}}=m_{3/2}M_{P}\ne0\,{\rm and}\,<\infty\label{eq:globala'}\\
{\rm b})\, M_{P}\rightarrow\infty & \frac{\Phi}{M_{P}}\ne0,\,\frac{Q}{M_{P}}\rightarrow0 & \frac{W}{M_{P}^{2}}\ne0\,{\rm and}\,<\infty\label{eq:globalb}
\end{eqnarray}
The first limit a) is the naive global limit. In this limit we recover
the global SUSY expressions $D_{i}W\rightarrow\partial_{i}W$ and
$V\rightarrow\partial_{i}WK^{i\bar{j}}\partial_{\bar{j}}\bar{W}$.
However in the presence of SUSY breaking there would be a cosmological
constant at the same scale as that of supersymmetry breaking. The
second and third however are the phenomenologically relevant limits
if we wish to incorporate the effects of SUSY breaking (i.e. generate
soft SUSY breaking terms in the low energy effective theory) and cancel
the cosmological constant that is generated when SUSY is broken. The
second is the limit that should be taken if one wishes to have a phenomenologically
relevant scenario like GMSB. However the gravitino mass goes to zero
in this limit. The third is relevant to all gravity/moduli mediated
scenarios and the gravitino mass is non-zero in the limit. Note that
in b) we've separated the chiral fields $\phi$ into {}``moduli''
$\Phi$ which typically take Planck scale vev's and matter fields
$Q$ which have either zero or small vev's (relative to the Planck
scale).

The first and second derivatives of the potential are given by the
following expressions (see for example \citep{Wess:1992cp}): 
\begin{eqnarray}
\partial_{i}V & = & e^{G}(G_{i}+G^{l}\nabla_{i}G_{l})+G_{i}V\label{eq:dV}\\
V_{i\bar{j}} & = & e^{G}(G_{i\bar{j}}+\nabla_{i}G_{k}\nabla_{\bar{j}}G^{k}-R_{i\bar{j}m\bar{n}}G^{m}G^{\bar{n}})+(G_{i\bar{j}}-G_{i}G_{j})V\label{eq:ddbarV}\\
V_{ij} & = & e^{G}(2\nabla_{i}G_{j}+G^{k}\nabla_{i}\nabla_{j}G_{k})+(\nabla_{i}G_{j}-G_{i}G_{j})V\label{eq:ddV}
\end{eqnarray}
$\nabla_{i}$ is the covariant derivative $\nabla_{i}X_{j}=\partial_{i}X_{j}-\Gamma_{ij}^{k}X_{k}$,
etc. The fermion mass matrix (ignoring mixing with the gravitino)
is 
\begin{equation}
m_{ij}=e^{G/2}(G_{ij}+G_{i}G_{j})=e^{K/2M_{P}^{2}}D_{i}D_{j}W,\label{eq:naivemass}
\end{equation}
where $D_{i}X_{j}=\nabla_{i}X_{j}+K_{i}X_{j}/M_{P}^{2}$. 

In the limits a) and a') above the fermion mass matrix goes to $m_{ij}\rightarrow\partial_{i}\partial_{j}W$
which is the usual expression in global SUSY. However as is well known
the fermions in a supergravity theory mix with the gravitino and after
unmixing the spin half mass matrix is actually \citep{Wess:1992cp}
\begin{eqnarray}
\tilde{m}_{ij} & =e^{G/2} & (\nabla_{i}G_{j}+\frac{1}{3}G_{i}G_{j})\label{fmass}\\
 & = & e^{K/2}e^{-i\phi_{W}}\left(D_{i}D_{j}W-\frac{2}{3}\frac{D_{i}WD_{j}W}{W}\right)\label{eq:fermmass}\\
 & \rightarrow & \left(W_{ij}-\frac{2}{3}\frac{\partial_{i}W\partial_{j}W}{W}\right).\nonumber 
\end{eqnarray}
The last line is valid for the global limits \eqref{eq:globala}\eqref{eq:globala'}.
Obviously if we had ignored gravity from the beginning this term would
not have been there - but what the above illustrates is that these
two theories i.e. the global theory and the limit of a local theory
(in the presence of SUSY breaking - are not necessarily the same.
Note that the second term goes to zero in the limit $W\rightarrow\infty$
(keeping $\partial W$ fixed) as it should. However in a limit where
the CC is zero, even if there are no large (i.e. Planck scale) moduli
fields, the second term is certainly not zero even in the naive decoupling
limit. For example if $W=c+\sigma_{i}\phi^{i}+\mu_{ij}\phi^{i}\phi^{j}+y_{ijk}\phi^{i}\phi^{j}\phi^{k},\, K=\sum_{i}|\phi|^{2}$,
$ $ the global theory would give a fermionic mass matrix $ $$m_{ij}=\mu_{ij}+O(\phi_{0})$
whereas even the naive limit \eqref{eq:globala} gives $\tilde{m}_{ij}=\mu_{ij}-2\sigma_{i}\sigma_{j}/3c+O(\phi_{0})$.
Note that if we wish also to tune the CC to zero we would need to
put $\sigma_{i}\sigma^{i}=3|c|^{2}/M_{P}^{2}$ which can only be satisfied
in the decoupling limit if we take the limits \eqref{eq:globala'}
\eqref{eq:globalb}. In this case we would have the second term in
the fermion mass matrix vanishing in the limit if we hold $\sigma$
fixed. On the other hand if $\sigma=O(M_{P})$ then again we would
 pick up a non-zero contribution from the mixing with the gravitino.
Of course the former limit corresponds to \eqref{eq:globala'} where
the gravitino mass is zero, while in the latter it is finite and non-zero
and is in fact the limit \eqref{eq:globalb}.

These issues are of relevance to arguments where the discussion of
supersymmetry breaking is primarily done within the global SUSY context.
For instance from the point of view of SUGRA, the arguments made in
\citep{Komargodski:2009jf} are actually dependent on the global limit
\eqref{eq:globala'} and also on having a flat Kaehler metric. To
understand the issues involved from a local perspective, let us consider
the full mass matrix
\begin{equation}
{\bf M}=\begin{pmatrix}M_{k\bar{l}}^{2}\, M_{kn}^{2}\\
M_{\bar{m}\bar{n}}^{2}\, M_{\bar{m}n}^{2}
\end{pmatrix}\label{eq:Mmatrix}
\end{equation}
The sub-matrices are given by (evaluating \eqref{eq:ddbarV}\eqref{eq:ddV}
at an extremum $dV=0$) 
\begin{eqnarray}
M_{\bar{l}k}^{2} & = & m_{\bar{l}}^{\,\, j}m_{jk}+\frac{1}{M_{P}^{2}}(K_{\bar{l}k}|F|^{2}-F_{\bar{l}}F_{k})-2|m_{3/2}|^{2}-R_{\bar{l}k\bar{m}n}F^{\bar{m}}F^{n},\label{eq:Mklbar}\\
M_{kn}^{2} & = & e^{K/2M_{P}^{2}}D_{n}D_{k}D_{i}WF^{i}-D_{k}D_{l}W\bar{m}_{3/2},\label{eq:Mkn}
\end{eqnarray}
where we've defined the gravitino mass function $m_{3/2}\equiv e^{K/2M_{P}^{2}}W/M_{P}^{2}$.
Now even in the global limit \eqref{eq:globala'}$ $
\begin{eqnarray*}
M_{\bar{l}k}^{2} & = & m_{\bar{l}}^{\,\, j}m_{jk}-R_{\bar{l}k\bar{m}n}F^{\bar{m}}F^{n},\\
M_{kn}^{2} & = & \nabla_{n}\nabla_{k}\nabla_{i}WF^{i}.
\end{eqnarray*}
Thus the statement that at tree level a zero eigenvector ${\bf v}$
of the mass matrix ${\bf m}=[m_{ij}]$ becomes a zero eigenvector
of the bosonic mass matrix ${\bf M}$ is not true in general even
in the limits \eqref{eq:globala}\eqref{eq:globala'}. It should be
stressed that non- trivial terms in the Kaehler potential (and hence
non-zero curvature) can be present even at tree level - for instance
from classically integrating out heavy states that couple to light
states. Thus the theorems proved in \citep{Komargodski:2009jf} are
strictly valid only in the global limit \eqref{eq:globala'} with
a flat Kaehler metric. For instance the existence of a flat direction
in O'R models (a so-called pseudo moduli space) depends in an essential
way on the assumption of no higher dimension operators (i.e. those
which are scaled by some mass which does not go to infinity with $M_{P}$)
in $W$or $K$. 

In the limit \eqref{eq:globalb} one needs to retain the second and
third terms of \eqref{eq:Mklbar} as well as the last term of \eqref{eq:Mkn}.
Needless to say the relationship between the fermionic zero mode and
the bosonic one is completely lost - even for a flat Kaehler metric
except in the limit \eqref{eq:globala'} - since there are terms proportional
to the gravitino mass which are absent in the fermion mass matrix.
In fact the unit vector ${\bf v}=[F^{i}/|F|^{2}]$ which defines the
Goldstino direction is not a zero mode of $[m_{ij}]$: $v^{i}m_{ij}=2\bar{m}_{3/2}v_{j}$
- after using $\partial_{i}V=0$. On the other hand 
\[
F^{i}\tilde{m}_{ij}=-\frac{2}{3}\frac{D_{i}W}{W}V|_{0}
\]
So the correct fermion mass matrix ($\tilde{m}$ the one which is
relevant in the presence of a gravitino \eqref{eq:fermmass}) has
a zero mode corresponding to $F^{i}/|F|^{2}$ provided that the CC
is tuned to zero. Again this has no simple relation to the bosonic
mass matrix.

The results proved in \citep{Komargodski:2009jf} have important consequence
in dynamical SUSY breaking in the context of GMSB theories. These
results are however valid only in the limit \eqref{eq:globala'} and
that too only provided that the Kaehler metric is flat.

\section{sGoldstino mass\label{sec:sGoldstino-mass}}

From \eqref{eq:dV}\eqref{eq:V} we see that at a extremum $\partial_{i}V=0$
with zero CC i.e. $V|_{0}=0$, we have $m_{ij}G^{i}=0$ so that the
Goldstino is the spinor $\chi_{g}^{i}=G^{i}\eta/3,\,\eta=G_{i}\chi_{g}^{i}$.
The sGoldstino is its complex scalar partner which therefore has averaged
squared mass (half the trace of the squared mass matrix, \citep{Gomez-Reino:2006dk}
(when $V|_{0}=0$) 
\begin{eqnarray}
M_{sg}^{2} & = & \frac{1}{3}V_{i\bar{j}}G^{i}G^{\bar{j}}\nonumber \\
 & = & \frac{1}{3}e^{G}(2G_{i\bar{j}}G^{i}G^{\bar{j}}-R_{i\bar{j}k\bar{l}}G^{i}G^{\bar{j}}G^{k}G^{\bar{l}})\nonumber \\
 & = & \frac{2}{3}K_{i\bar{j}}\frac{F^{i}F^{\bar{j}}}{M_{P}^{2}}-\frac{1}{|F|^{2}}R_{i\bar{j}k\bar{l}}F^{i}F^{\bar{j}}F^{k}F^{\bar{l}}).\label{sGmass}\\
 & = & 2m_{3/2}^{2}-\frac{1}{3M_{P}^{2}m_{3/2}^{2}}R_{i\bar{j}k\bar{l}}F^{i}F^{\bar{j}}F^{k}F^{\bar{l}}\label{sgmass2}
\end{eqnarray}
In the last equality above we've imposed $V|_{0}=0$. This formula
immediately tells us that (since $F\sim M_{P}m_{3/2}$ when the cosmological
constant is cancelled), unless the curvature on moduli space is enhanced
way above the Planck scale or there is cancellation between the two
terms, the sGoldstino mass is of order of the gravitino mass. 

Several special cases are of some interest:
\begin{enumerate}
\item In any model in which the metric on moduli space is flat (as is the
case with most dynamical SUSY breaking models) the (mean) sGoldstino
mass is $\sqrt{2}m_{3/2}$.
\item Einstein Kaehler space $R_{i\bar{j}k\bar{l}}=\sigma K_{i\bar{j}}K_{k\bar{l}}$.
In this case $m_{sg}^{2}=(2-3\sigma)m_{3/2}^{2}$. The no-scale model
for instance corresponds to $\sigma=2/3$.
\end{enumerate}
Notice that in the global limit \eqref{eq:globala'} the mean sGoldstino
mass is actually zero in the flat Kaehler metric case. Writing $m_{3/2}M_{P}=\mu^{2}<\infty$,
as $M_{P}\rightarrow\infty$, we have \eqref{sgmass2}
\begin{equation}
M_{sg}^{2}=-\frac{1}{\mu^{4}}R_{i\bar{j}k\bar{l}}F^{i}F^{\bar{j}}F^{k}F^{\bar{l}}.\label{eq:globalSGmass}
\end{equation}
This means that unless the RHS is positive and non-zero one or other
of the sGoldstinos is tachyonic. In the case that the metric is flat,
if we also impose the condition of no tachyons then clearly there
is locally a (complex) flat direction and both sGoldstinos have zero
mass. Thus any sensible theory in the limit \eqref{eq:globala'} must
necessarily have a non-zero curvature in field space. 

An example of a case where the sGoldstino mass is enhanced above the
gravitino mass is the Kitano model \citep{Kitano:2006wz} in which
\begin{equation}
K=S\bar{S}-\frac{(S\bar{S})^{2}}{\Lambda^{2}}+q\bar{q}+\tilde{q}\bar{\tilde{q}}+\frac{\lambda^{2}}{(4\pi)^{2}}S\bar{S}\ln\frac{(S\bar{S})^{2}}{\Lambda'^{2}},\, W=c+\mu^{2}S+\lambda Sq\tilde{q}.\label{eq:Kitano}
\end{equation}
Here $(q,\tilde{q})$ are messengers, $S$ is some gauge neutral state
(for example a modulus of string theory and $\Lambda$ is the scale
at which for instance Kaluza-Klein states or string states have been
integrated out. The last term in the Kaehler potential is a messenger
one loop effect. This gives $R_{S\bar{S}S\bar{S}}=-4/\Lambda^{2}$.
$S$ defines the direction of SUSY breaking . In this case 
\begin{equation}
M_{sg}^{2}=<M_{S}^{2}>=2m_{3/2}^{2}(1+6\frac{M_{P}^{2}}{\Lambda^{2}}(1+6(\frac{\lambda}{4\pi})^{4}\frac{M_{P}^{2}}{\Lambda^{2}})\simeq12\left(\frac{M_{P}m_{3/2}}{\Lambda}\right)^{2}.\label{eq:Msg}
\end{equation}
The last approximation follows from $\Lambda<M_{P}$ and the stability
condition for the $S$ mass squared matrix which implies that $\lambda/4\pi<\Lambda/M_{P}$,
(see \citep{Kitano:2006wz}) which is also the condition that the
expansion in messenger loops makes sense. It would appear that by
choosing the cutoff sufficiently low one can get a sGoldstino mass
well above the gravitino mass scale. However this not correct. There
are two stability conditions to satisfy at the minimum ($S=\Lambda^{2}/\sqrt{12}M_{P},\, q=\tilde{q}=0$)
- one coming from the mass matrix for $S$ (which should include the
contribution coming from messenger loops), and the other from that
for the messengers. After tuning the CC to zero (giving $\mu^{2}\simeq\sqrt{3}m_{3/2}M_{P}$)
one gets the following window for the messenger coupling:
\begin{equation}
\frac{12\sqrt{3}}{(4\pi)}\frac{m_{3/2}M_{P}^{3}}{\Lambda^{4}}<\frac{\lambda}{4\pi}<\frac{\Lambda}{M_{P}}.\label{eq:Window}
\end{equation}
Furthermore the gaugino mass is given by \citep{Giudice:1998bp} 
\begin{equation}
M\simeq\frac{\alpha}{4\pi}\frac{F^{S}}{S}\simeq\frac{\alpha}{4\pi}\frac{\sqrt{3}m_{3/2}M_{P}}{\lambda\Lambda^{2}/(2\sqrt{3}M_{P})}\label{eq:gauginomass}
\end{equation}
\uline{I}n order to have an open window \eqref{eq:Window} the
gravitino mass is bounded below and in order to have GMSB dominance
it needs to be much smaller than the gaugino mass. For gaugino masses
(and other soft parameters) at the weak scale (i.e. $M\sim100GeV$)
this implies,
\begin{equation}
10^{-5}GeV\leq m_{3/2}\ll100GeV.\label{eq: gravitinobound}
\end{equation}
From \eqref{eq:Msg} and \eqref{eq:gauginomass} we also have
\begin{equation}
M_{S}=2\sqrt{3}m_{3/2}\left(\frac{4\pi\lambda M}{6\alpha m_{3/2}}\right)^{1/2}\sim10\sqrt{\lambda}\sqrt{m_{3/2}M},\label{eq:mX2}
\end{equation}
where in the last relation we have used $\alpha/4\pi\sim10^{-2}$
for the gauge coupling. Thus the modulus mass is constrained by the
scale of soft masses - it is in fact well below that scale. Note that
the above constraints \eqref{eq: gravitinobound} for the gravitino
mass as well as the corresponding value for the sGoldstino mass, are
incompatible with standard cosmology (for a recent discussion see
\citep{Feng:2010ij}). 

Let us see what happens in general O'Raifeartaigh (O'R) type models
for SUSY breaking embedded in SUGRA. We take the following potentials
:
\begin{eqnarray}
K & = & S\bar{S}-\frac{(S\bar{S})^{2}}{\Lambda^{2}}+\sum_{i}q^{i}\bar{q^{i}}+\frac{1}{(4\pi)^{2}}tr|{\cal M}|^{2}\ln\frac{|{\cal M}|^{2}}{\Lambda'^{2}},\label{eq:Kgeneral}\\
W & = & c+\mu^{2}S+{\cal M}_{ij}(S)q^{i}q^{j}\label{eq:Wgeneral}
\end{eqnarray}
where 
\begin{equation}
{\cal M}_{ij}=\lambda_{ij}S+m_{ij}.\label{eq:massmatrix}
\end{equation}
The superpotential here is written in the so-called canonical form
- every O'R model can be rewritten in this form (see for example \citep{Komargodski:2009jf}).
Then \eqref{eq:gauginomass} will be replaced by 
\begin{equation}
M\simeq\frac{\alpha}{4\pi}F^{S}\partial_{S}\ln\det{\cal M}.\label{eq:gaugino2}
\end{equation}
In this case the cutoff $\Lambda$ can in principle be decoupled from
the soft mass scale since one could take (roughly speaking) $\lambda^{-1}m\gg S_{0}\sim\Lambda^{2}/M_{P}$.
In this case one can in fact have a standard cosmological scenario.
Nevertheless the stability constraints impose restrictions which are
rather unnatural. Schematically these constraints now take the form,
\begin{equation}
\frac{m_{3/2}M_{P}}{|\lambda^{-1}{\cal M}|^{2}}<\lambda<4\pi\frac{|\lambda^{-1}{\cal M}|}{\Lambda}\label{eq:window2}
\end{equation}
where $|{\cal M}|$ stands for the scale of the messenger mass matrix.
This (together with \eqref{eq:gaugino2} and $M\sim100GeV$) gives
only the lower bound $m_{3/2}>0.1eV$ and so is compatible with the
standard cosmological scenario. However in this case we also have
the bounds 
\begin{equation}
\Lambda<10^{9}GeV,\,|m|<10^{4}GeV\label{eq:Lamdambounds}
\end{equation}
The bound on $m$ is of course quite unnatural for a SUSY preserving
term. Secondly the cutoff $\Lambda$ is expected to be some physical
scale such as the Kaluza-Klein scale of string theory. Thus this bound
on $\Lambda$ would imply a KK scale which is unnaturally low. In
fact it is often the case in the GMSB literature, that the cut off
is assumed to be at the GUT scale! Thus it appears that the only way
we can get a low enough gravitino mass (as well as a high sGoldstino
mass) in these theories is by making two unnatural choices of mass
parameters.$ $

\[
\]

\section{R-axion and SUSY breaking\label{sec:R-symmetry-and-SUSY}}

\subsection{Global SUSY bound}

Let us first recapitulate the argument of \citep{Dine:2009sw} within
the global supersymmetric context. Let the theory have a R-symmetry
generated by the Killing vector $k^{i}(\Phi)$. Here $i$ labels the
chiral superfields of the theory. For a linearly realized R-symmetry
$k^{i}=iq^{i}\Phi^{i}\,({\rm no\, sum})$ where $q^{i}$ is the R-charge
of $\Phi^{i}$. Since under an R-symmetry the superpotential transforms,
we have the Killing equation (using the notation $\partial_{i}\equiv\partial/\partial\Phi^{i})$
\begin{equation}
k^{i}W_{i}=i2W,\label{eq:R}
\end{equation}
where as usual we've taken the charge of $W$ to be two. Let the Kaehler
potential of the theory be $K(\Phi,\bar{\Phi})$ so that the Kaehler
metric is $K_{i\bar{j}}$. For any pair of vectors ${\bf U}=\{U^{i}\},{\bf V}=\{V^{i}\}$,
we define the inner product $<{\bf U},{\bf V}>=\overline{<{\bf V},{\bf U}>}\equiv K_{i\bar{j}}V^{i}\bar{U}^{\bar{j}}$.
Then putting $\bar{U}^{\bar{j}}=K^{\bar{j}l}\partial_{l}W$and $V^{i}=k^{i}$
the Killing equation \eqref{eq:R} becomes
\begin{equation}
<{\bf U},{\bf V}>=i2W\label{eq:R2}
\end{equation}
Let the scalar component of $\Phi^{i}=\phi^{i}$. If one or more of
the charged fields acquire a non-zero vacuum expectation value $\phi_{0}^{i}$,
the R-symmetry is spontaneously broken and we may isolate the axion
field by writing $\phi^{i}=\phi_{0}^{i}e^{iq^{i}a(x)}$. The kinetic
term for the R-axion then becomes 
\[
K_{i\bar{j}}\partial\Phi^{i}\partial\bar{\Phi}^{\bar{j}}\rightarrow K_{i\bar{j}}k_{0}^{i}k_{0}^{\bar{j}}(\partial a)^{2}
\]
so that the axion decay constant $f_{a}$ is given by (here and in
what follows the subscript $0$ indicates evaluation at the minimum
of the potential), 
\[
f_{a}^{2}=K_{i\bar{j}}k_{0}^{i}k_{0}^{\bar{j}}=<{\bf V}_{0},{\bf V}_{0}>.
\]
 Then using the Cauchy-Schwarz inequality and equation \eqref{eq:R2}
we have,
\begin{equation}
4|W_{0}|^{2}=|<{\bf U}_{0},{\bf V}_{0}>|^{2}\leq<{\bf U}_{0},{\bf U}_{0}><{\bf V}_{0},{\bf V}_{0}>=|F|_{0}^{2}f_{a}^{2},\label{eq:inequality}
\end{equation}
where $|F|^{2}=K^{i\bar{j}}\partial_{i}W\partial_{\bar{j}}\bar{W}$.

\subsection{SUGRA bound}

\[
\]
The above inequality is valid only in global SUSY. As we pointed out
earlier the superpotential has no meaning in and of itself in global
supersymmetry. Only the derivatives of the superpotential have physical
significance. Thus the formula \eqref{eq:inequality} makes sense
only in the context of supergravity. However supergravity is invariant
under the Kaehler transformations $K\rightarrow K+\Lambda+\bar{\Lambda},\, W\rightarrow e^{-\Lambda}W$.
The above formula is not invariant under these transformations and
hence is not valid as it stands in supergravity. Let us therefore
work out the Kaehler invariant form of the above discussion. 

Since the Kaehler potential is invariant under the R-symmetry we have
\begin{equation}
k^{i}\partial_{i}K+\bar{k}^{\bar{i}}\partial_{\bar{i}}K=iq^{i}\phi^{i}\partial_{i}K-iq^{i}\bar{\phi}^{\bar{i}}\partial_{\bar{i}}K=0\Rightarrow q^{i}\phi^{i}\partial_{i}K=\overline{q^{i}\phi^{i}\partial_{i}K}.\label{eq:kK}
\end{equation}
In supergravity the order parameter for SUSY breaking, i.e. the F-term
(for the moment we ignore gauge interactions) is given by the expression
\begin{equation}
F_{i}=e^{K/2}D_{i}W\rightarrow e^{(\bar{\Lambda}-\Lambda)/2}F_{i}\Rightarrow|F_{i}|\rightarrow|F_{i}|\label{eq:F}
\end{equation}
where the second pair of relation indicate the Kaehler transformation
properties and we've set $M_{P}=1$. 

If the theory has an R-symmetry then the superpotential satisfies
equation \eqref{eq:R}. So we get 
\[
k^{i}F_{i}=(2i+k^{i}K_{i})e^{K/2}W=i(2+\sum_{i}q^{i}\phi^{i}K_{i})e^{K/2}W
\]
Evaluating this at the minimum of the potential and identifying the
gravitino mass as $m_{3/2}=e^{K/2}|W|_{0}$, we have 
\[
|<{\bf k},{\bf F}>_{0}|=|2+\sum_{i}q^{i}\phi_{0}^{i}\partial_{i}K_{0}|m_{3/2}.
\]
Using the Cauchy-Schwarz inequality we get (again putting $f_{a}^{2}=<{\bf k}_{0},{\bf k}_{0}>$)
instead of \eqref{eq:inequality}, the following;

\begin{equation}
f_{a}^{2}<{\bf F},{\bf F}>|_{0}\geq|2+\sum_{i}q^{i}\phi_{0}^{i}\partial_{i}K_{0}|^{2}m_{3/2}^{2}.\label{eq:inequality2}
\end{equation}
Note however that this is still not Kaehler invariant! The reason
is that the R-symmetry equation is not Kaehler covariant. However
if we assume that the theory is such that the sum on the right hand
side is positive then we can write, (restoring $M_{P}$) 
\begin{equation}
f_{a}^{2}<{\bf F},{\bf F}>_{0}\geq4m_{3/2}^{2}M_{P}^{4}.\label{eq:inequality3}
\end{equation}
However as we've observed this is model dependent. Nevertheless it
is the obvious Kaehler invariant generalization of \eqref{eq:inequality}.
It should be observed however that the requirement of setting the
cosmological constant (CC) to zero implies that this is actually an
inequality for the axion decay constant. Since we must fine tune the
parameters of the theory such that 
\begin{equation}
<{\bf F},{\bf F}>_{0}\equiv F_{i}K^{i\bar{j}}\bar{F}_{\bar{j}}|_{0}=3m_{3/2}^{2}M_{P}^{2},\label{eq:CC}
\end{equation}
 we get from \eqref{eq:inequality3},
\begin{equation}
f_{a}^{2}\geq\frac{4}{3}M_{P}^{2}\label{eq:faineq}
\end{equation}
Actually it is clear from \eqref{eq:inequality2} and \eqref{eq:CC}
that, since for a generic theory the first factor on the RHS of that
inequality i.e. the expression $ $$4M_{P}^{4}|1+\frac{1}{2M_{P}^{2}}\sum_{i}q^{i}\phi_{0}^{i}\partial_{i}K_{0}|^{2}$
is Planck scale, we will have a Planck scale axion decay constant,
\[
f_{a}^{2}\gtrsim M_{P}^{2},
\]
and this is of course both Kaehler invariant and valid for generic
theories with a spontaneously broken R-symmetry.

\section{Global limits of a local SUSY model with FI terms\label{sec:FI-terms-in}}

\subsection{Classical issues}

Let us now discuss the issue of Fayet-Illiopoulos terms from the point
of view of the different global limits. In \citep{Komargodski:2009pc}
certain problems with extending global theories with FI terms to SUGRA
were discussed (see also \citep{Komargodski:2010rb}\citep{Dienes:2009td}).
In \citep{deAlwis:2012tp} a different approach is taken. Here we
discuss some aspects of \citep{deAlwis:2012tp} that are relevant
to the issues that we've addressed in this note. There is of course
no conflict between the statements made there (or in this note) and
the arguments of \citep{Komargodski:2009pc} since the class of SUGRA
theories that we consider do not have a well defined global limit.

We begin by introducing a SUGRA theory%
\footnote{This model along with a discussion of its global limits was circulated
amongst a few workers in the field in October 2010. Since then a similar
model has been published by other authors \citep{Catino:2011mu}. %
} with an FI term that is manifestly consistent, at least at the classical
level %
\footnote{Quantum anomalies can be cancelled by adding additional fields and
or Green-Schwarz terms as we shall see.%
}. The model is given by

\begin{equation}
G\equiv K+\hat{\xi}V+M_{P}^{2}\ln\frac{|W|^{2}}{M_{P}^{6}}\label{eq:Gmodel}
\end{equation}
with 
\begin{equation}
K=\bar{S}e^{V}S+\sum\bar{\Phi}\Phi,\, W=\left(\frac{S}{M_{P}}\right)^{\hat{\xi}/M_{P}^{2}}W_{I}(\Phi)\label{eq:KWmodel}
\end{equation}
In the simplest version discussed $W_{I}$ is taken to be a constant.
However in general we can take it to be a gauge invariant holomorphic
function involving the other chiral fields $\Phi$ as well as possibly
$S$ itself. The superfield $G$ is then invariant under the gauge
transformations
\begin{eqnarray}
V & \rightarrow & V+i(\Lambda-\bar{\Lambda})\label{eq:Lgauge}\\
S & \rightarrow & e^{-i\Lambda}S\label{eq:Sgauge}
\end{eqnarray}
with the other fields transforming appropriately. Since by well-known
arguments off-shell supergravity can be expressed entirely in terms
of the function $G$ (and the holomorphic gauge coupling function
$f(\Phi)$) this model gives a gauge invariant supergravity with an
FI term. 

Now the global theory is well defined even with the addition of an
FI term $\xi V$ to the invariant Kaehler potential $K$; under the
gauge transformation $V\rightarrow V+i\Lambda-i\bar{\Lambda}$, this
term is invariant because of the $\int d^{4}\theta$ integral. In
SUGRA however the Kaehler potential comes in an exponential and will
not give an invariant theory unless one adds a harmonic (i.e. chiral
plus anti-chiral) piece that transforms in such a fashion as to cancel
the gauge variance of the $\xi V$ term. So after a Kähler transformation
the full superspace integral is taken to be (with ${\bf E}$ being
the full superspace supervielbein determinant) 
\begin{eqnarray*}
-3M_{P}^{2}\int d^{4}\theta{\bf E}e^{-[K+\hat{\xi}V+\hat{\xi}(\ln(S/M_{P})+h.c.)]/3M_{P}^{2}}.
\end{eqnarray*}
Now in this expression, let us take the global limit 
\begin{eqnarray}
M_{P}^{2} & \rightarrow\infty\,, & \mbox{E}\rightarrow1,\label{eq:globallimit}\\
\xi & =\frac{\hat{\xi}}{M_{P}^{2}} & \rightarrow O(\frac{1}{M_{P}^{2}}).\label{eq:xihatfixed}
\end{eqnarray}
 The last relation implies keeping $\hat{\xi}$ fixed as we take the
limit. Then we get 
\begin{equation}
\int d^{4}\theta[-3M_{P}^{2}+(K+\hat{\xi}V+\hat{\xi}(\ln S/M_{P}+h.c.)+O(\frac{1}{M_{P}^{2}})]\rightarrow\int d^{4}\theta(K+\hat{\xi}V)\label{eq:limit}
\end{equation}
In the last expression we have used the chirality of $\ln S$ (i.e.
$D_{\alpha}\ln\bar{S}=D_{\alpha}\bar{S}/\bar{S}=0$$ $). Note that
the apparent singularity of the formalism is irrelevant since all
that means is (see for example the discussion of the potential in
the last section) that for $\xi<1$ there is no gauge invariant $S=0$
minimum of the potential.

On the other hand if $\xi\ne0$ is fixed in the limit $M_{P}^{2}\rightarrow\infty$
(whether or not $\hat{\xi}$ is quantized in Planck units) we get
in the limit \eqref{eq:globallimit} instead of \eqref{eq:limit}
the expression,
\[
\int d^{4}\theta(K+\xi M_{P}^{2}(V+\ldots)
\]
which means that the limit does not exist!

It is in fact instructive to study the case $\xi<1$ in a little more
detail in the simple model where $W_{I}$ is independent of $S$ and
is just a function of a neutral field $\Phi$. Note that in this model
$K=\bar{S}e^{V}S+\bar{\Phi}\Phi$. The potential in the global limit
\eqref{eq:globala} (with $\hat{\xi}$ fixed) is 
\begin{equation}
V_{global}=\mid\frac{\partial W_{I}}{\partial\Phi}\mid^{2}+\frac{g^{2}}{8}(S\bar{S}+\hat{\xi})^{2},\label{eq:Vglobal}
\end{equation}
and there is a gauge invariant ground state $S=0$ regardless of the
value of $\hat{\xi}$ ($>0$). Note also that at this minimum supersymmetry
is broken and (assuming there is a solution to $\partial W_{I}/\partial\Phi=0$)
\[
V_{global,0}=\frac{g^{2}}{8}\hat{\xi}^{2}
\]
 $ $ On the other hand before taking the limit the potential of the
theory is
\begin{eqnarray}
V=e^{(S\bar{S}+\Phi\bar{\Phi})/M_{P}^{2}} & \left(\frac{S\bar{S}}{M_{P}^{2}}\right)^{\hat{\xi}/M_{P}^{2}} & [\frac{\mid W_{I}\mid^{2}}{S\bar{S}}(\frac{\hat{\xi}}{M_{P}^{2}}+\frac{S\bar{S}}{M_{P}^{2}})^{2}+\mid\partial_{\Phi}W_{I}+\frac{\bar{\Phi}}{M_{P}^{2}}W_{I}\mid^{2}-3\frac{\mid W_{I}\mid^{2}}{M_{P}^{2}}]\nonumber \\
 &  & +\frac{g^{2}}{8}(S\bar{S}+\hat{\xi})^{2}\label{eq:Vsugra}
\end{eqnarray}
Now the global limit in which $M_{P}^{2}\rightarrow\infty$ with $W_{I},\hat{\xi},S,\Phi$
fixed gives us back \eqref{eq:Vglobal}. On the other hand in the
limit a') (see \eqref{eq:globala'}) (with $W_{I}/M_{P}\equiv\mu^{2}$
fixed but without assuming that $s_{0}\rightarrow O(1/M_{P}^{2})$),
we have
\begin{equation}
V\rightarrow e^{S\bar{S}/M_{P}^{2}}[\mu^{4}\frac{S\bar{S}}{M_{P}^{2}}+|\partial_{\phi}W_{I}|^{2}-\mu^{4}]+\frac{g^{2}}{8}(S\bar{S}+\hat{\xi})^{2},\label{eq:global'}
\end{equation}
which again has a minimum at $S=0$. However as can be seen from \eqref{eq:Vsugra},
for $\xi=\hat{\xi}/M_{P}^{2}<1$, the minimum of the potential \eqref{eq:Vsugra}
is at (for finite $M_{P}$) 
\[
\frac{S_{0}\bar{S}_{0}}{M_{P}^{2}}\ne0.
\]
Thus in this case the existence of the gauge invariant minimum is
an artifact of the decoupling limit. 

The model is easily generalized to the case when the superpotential
of the global theory ($W_{I}$) is not independent of the field $S$.
In this case we need at least one other field which is charged (with
charge $-1$). The invariant superpotential is 
\begin{equation}
W_{I}=W_{I}(S,\Phi),\label{eq:WISP}
\end{equation}
where $\Phi$ now stands for a set of fields which must have at least
one charged field. The invariant Kaehler potential plus FI term is
taken as before to be 
\begin{equation}
\hat{K}=K+\hat{\xi}V+\hat{\xi}\ln(S\bar{S}/M_{P}^{2}).\label{eq:Khat}
\end{equation}
Note that there is now an ambiguity in the choice of the last term.
For instance if there is only one other charged field in the set $\Phi$
(say $\tilde{S}$ with charge $-1$) then an equally valid extension
to SUGRA of the original global theory would have instead $ $$\hat{K}=K+\hat{\xi}V-\hat{\xi}\ln(\tilde{S}\bar{\tilde{S}})/M_{P}^{2})$.
This means that there will be several possible SUGRA extensions of
a given global theory. In any case the point is that (taking for comparison
with the previous discussion the extension (\ref{eq:Khat}), we will
essentially have the same potential as \eqref{eq:Vsugra} with some
simple modifications. Namely we have 
\begin{eqnarray}
V=e^{(S\bar{S}+\sum\Phi\bar{\Phi})/M_{P}^{2}} & \left(\frac{S\bar{S}}{M_{P}^{2}}\right)^{\hat{\xi}/M_{P}^{2}} & [\frac{\mid W_{I}\mid^{2}}{S\bar{S}}\mid\frac{\hat{\xi}}{M_{P}^{2}}+\frac{S\bar{S}}{M_{P}^{2}}+\frac{\partial\ln W_{I}}{\partial\ln S})\mid^{2}+\mid\partial_{\Phi}W_{I}+\frac{\bar{\Phi}}{M_{P}^{2}}W_{I}\mid^{2}\nonumber \\
 &  & -3\frac{\mid W_{I}\mid^{2}}{M_{P}^{2}}]+\frac{g^{2}}{8}(S\bar{S}+\sum q_{\Phi}|\Phi|^{2}+\hat{\xi})^{2}.\label{eq:Vsugra2}
\end{eqnarray}
Finding the minima of this potential is of course much more complicated.
However it is clear that as in the simpler case analyzed before, for
$\hat{\xi}<M_{P}^{2}$ the potential $V\rightarrow+\infty$ for both
limits $S\bar{S}\rightarrow0,\,+\infty$, and so the minimum will
be at $|S|_{0}\ne0$. Thus quite generally it is the case that when
$0<\xi<1$ the minimum of the potential will not be $U(1)$ symmetric
except in the global limit. Generically supersymmetry will also be
broken. 

Finally we should stress that even if it were the case that the gauge
group is compact so that $\hat{\xi}$ is quantized in Planck units
\citep{Seiberg:2010qd}\citep{Distler:2010zg}\citep{Hellerman:2010fv},
the theory can exist as a valid effective field theory if the scale
of the superpotential $W_{I}$ is chosen to be well below the Planck
scale.

\subsection{Quantum issues}

So far the discussion of this model has been entirely classical. In
fact at the quantum level there is an anomaly in the model, which
can be dealt with either by adding extra fields (as in \citep{Catino:2011mu})
or by including a Green-Schwarz anomaly canceling sector.

Let us consider extending the model so that the $U(1)$ anomaly is
cancelled by the Green-Schwarz mechanism. So we take the model of
\eqref{eq:KWmodel} (for simplicity without the $\Phi$ field) and
add a field $T$ with the non-linear $U(1)$ transformation rule 
\begin{equation}
T\rightarrow T-iM\Lambda,\label{eq:Ttrans}
\end{equation}
and an additional (gauge invariant) Kaehler potential%
\footnote{Note this is to be contrasted with the case of a $\Delta K=(T+\bar{T}+MV)^{2}$
where by a redefinition $T=T'-\xi/2M$ the FI term can be removed
as discussed in \citep{Dienes:2009td,Catino:2011mu}.%
}
\begin{equation}
\Delta K=-3\ln(T+\bar{T}+MV).\label{eq:DeltaK}
\end{equation}
Note that again we've set $M_{P}=1$. To incorporate anomaly cancellation
we now take the gauge coupling function to be
\begin{equation}
f=\frac{1}{g^{2}}+b_{UU}T,\label{eq:f}
\end{equation}
where $b_{UU}$ is the pure gauge anomaly coefficient. In addition
we will need to add a curvature squared term with a coefficient $b_{RR}T$
(where $b_{RR}$ is the mixed gauge gravitational anomaly coefficient)
to cancel the mixed anomaly.

The potential is positive definite because of the no-scale form of
$\Delta K$; 
\begin{eqnarray*}
V & = & V_{F}+V_{D},\\
 & = & \frac{\mid W_{I}\mid^{2}}{(T+\bar{T})^{3}}e^{S\bar{S}}(S\bar{S})^{\xi-1}(S\bar{S}+\xi)^{2}\\
 &  & +\frac{g^{2}}{8(1+\frac{g^{2}}{2}b_{UU}(T+\bar{T}))}(S\bar{S}+\xi-\frac{3M}{T+\bar{T}})^{2}.
\end{eqnarray*}
For $\xi>1$ it has a SUSY minimum at $S\bar{S}=0,\,\Re T=3M/2\xi$
with $V_{0}=0$. Note that a non-zero value of $\xi$ is crucial for
avoiding runaway behavior in the $\Re T$ direction. On the other
hand for $\xi\leq1$ $\Re T$ runs away to infinity at the global
minimum. There is however a local minimum where SUSY and gauge invariance
are broken if the condition for a certain quartic in $\Re T$ to have
real roots, is satisfied by the coefficients $M,\xi$.

\section{Conclusions}

In this work we have elaborated on the well known phenomenon that
a generic supergravity theory can have different global limits. The
usual action of global supersymmetry is simply one such limit. If
one starts from that limit to build supergravity only a limited class
will emerge. The point is when supersymmetry is broken the relevant
limit which would lead to a zero CC (or at least one which is parametrically
smaller than the SUSY breaking scale) is not the one which would have
been obtained as a global supersymmetric theory where gravity was
not taken into account. We've pointed out in this paper that issues
related to the scalar partner of the goldstino (which has the potential
to cause cosmological moduli problems) can be meaningfully addressed
only within the SUGRA context. Also we show that a bound on the superpotential
which may be derived in the global SUSY context disappears when it
is rederived in the context of SUGRA and the cosmological constant
is tuned to zero. Finally we addressed issues relating to Fayet-Illiopoulos
terms in SUGRA and global SUSY from the perspective of these different
limits.

\section{Acknowledgments}

I wish to thank Michael Dine, Oliver De Wolfe, Fernando Quevedo and
Marco Serone for discussions. I also wish to thank Keith Dienes and
Brooks Thomas for collaboration on \citep{deAlwis:2012tp}, agreeing
to let me use some of that material in section 5 of this work, and
for comments on the manuscript. The award of an SFB fellowship from
the University of Hamburg and DESY, and a visiting professorship at
the Abdus Salam ICTP are gratefully acknowledged. This research is
partially supported by the United States Department of Energy under
grant DE-FG02-91-ER-40672. 

\[
\]
 \bibliographystyle{apsrev}
\bibliography{myrefs}

\end{document}